\documentclass[12pt,preprint]{aastex}

\begin{document}
\slugcomment{Submitted to ApJ Letter}

\title{An Ultraluminous Supersoft X-ray Source in M81: An Intermediate-Mass Black Hole?}

\author{Jifeng Liu and Rosanne Di Stefano}

\affil{Harvard-Smithsonian Center for Astrophysics}

\begin{abstract}

Ultraluminous supersoft X-ray sources (ULSSS) exhibit supersoft spectra with blackbody temperatures of 50-100 eV and bolometric luminosities above $10^{39}$ erg s$^{-1}$, and are possibly intermediate mass black holes (IMBHs) of $\ge10^3 M_\odot$ or massive white dwarfs that are progenitors of type Ia supernovae.  In this letter we report our optical studies of such a source in M81, M81-ULS1, with HST archive observations.  M81-ULS1 is identified with a point-like object, the spectral energy distribution of which reveals a blue component in addition to the companion of an AGB star.  The blue component is consistent with the power-law as expected from the geometrically-thin accretion disk around an IMBH accretor, but inconsistent with the power-law as expected from the X-ray irradiated flared accretion disk around a white dwarf accretor.  This result is strong evidence that M81-ULS1 is an IMBH instead of a white dwarf.

\end{abstract}


\section{INTRODUCTION}

Luminous supersoft X-ray sources were discovered with the Einstein Observatory
and were established as an important new class of X-ray binaries based on ROSAT
observations of 18 sources in the Milky Way and the Magellanic Clouds (Kahabka
\& van den Heuvel 1997 and reference therein). 
These sources have extremely soft spectra with equivalent blackbody
temperatures of 15-80 eV and are highly luminous with bolometric luminosities
of $10^{35} - 10^{38}$ erg s$^{-1}$. 
They are thought to be white dwarfs that are steadily or cyclically burning
hydrogen accreted onto the surface. Remarkably, the accretion rates must be in
a narrow range of $1-4\times10^{-7} M_\odot yr^{-1}$, and the resultant
luminosities are below $10^{39}$ erg s$^{-1}$ as observed.
White dwarfs with steady nuclear burning are promising Type Ia supernova
progenitors because, unlike explosive nova events, they can retain the accreted
matter and their mass can increase until it approaches the Chandrasekhar limit.

More and more supersoft sources have been discovered in distant galaxies with
the advent of the Chandra and XMM-Newton X-ray observatories.
These distant supersoft sources, as compared to the canonical ones, are
generally hotter, more luminous and often associated with spiral arms,
suggesting that some are young and massive systems (Di Stefano \& Kong 2004).
Some have bolometric luminosities, as derived from blackbody models, far above
the Eddington luminosity ($L_{Edd} = 4\pi G M m_p c/\sigma_T \approx
1.3\times10^{38} (M/M_\odot)$ erg s$^{-1}$) for white dwarfs, which we call
ultraluminous supersoft sources (ULSSS), or ULS as a more compact name.
One prototype ULSSS is M101-ULX1, which showed supersoft spectra of $50\sim100$
eV,  0.3-7 keV luminosities of $3\times10^{40}$ erg s$^{-1}$ and  bolometric
luminosities of $\sim 10^{41}$ erg s$^{-1}$ during its 2004 outburst (Kong et
al. 2004). 
While the supersoft spectrum can be explained by a white dwarf burning accreted
materials on its surface, the white dwarf can not explain the luminosities that
are extremely super-Eddington for white dwarfs of $\leq 1.4 M_\odot$.
On the other hand, an intermediate mass black hole (IMBH) of $\gtrsim 10^3
M_\odot$, as Kong et al. suggested, can naturally explain both the high
bolometric luminosities and the supersoft spectrum given the scaling relation
$T_{in} \propto M^{-1/4}$ between the black hole mass and the accretion disk
inner edge temperature.
%
%

Intermediate mass black holes ($10^2$-$10^5M_\odot$) have played an important
role in the hierarchical merging scenario of galaxy formation (Madau \& Rees
2001).
They have long been searched for in the cores of globular clusters, but
iron-clad evidence is lacking (for a review, see Miller \& Colbert 2004).
Recently, much interest has been directed toward ultraluminous X-ray sources
(ULXs), which have been found to exist outside the nuclear regions with
$L_X\sim10^{39}$-$10^{41}$ erg/sec in many nearby galaxies (Miller \& Colbert
2004).
Such luminosities require IMBHs of $\ge10^3 M_\odot$ if ULXs radiate at
$10^{-2}$ Eddington level as do many active galactic nuclei and Galactic X-ray
binaries.
One type of evidence for IMBHs comes from the X-ray spectroscopy that reveals a
cool accretion disk component with the inner edge temperature of a few $\times
10$ eV, which corresponds to an IMBH of a few $\times 10^3 M_\odot$ if the
inner edge of the accretion disk corresponds to the last stable orbit (Miller
et al. 2004).
However, in many cases the cool component is dominated by a hard power-law
component.  This hard component resembles those from the hot corona in stellar
black hole X-ray binaries that will erode and envelope the inner part of the
accretion disk, raising doubts against the IMBH interpretation and suggestions
that ULXs may be black holes of 20-100 $M_\odot$ in very high state (Done \&
Kubota 2006; Soria 2007).
Indeed, ULXs can, from a theoretical perspective, be stellar mass black holes
with beaming effects or radiating at super-Eddington levels with the
photon-bubble instability operating in a radiation pressure dominated
magnetized accretion disk (Begelman 2002).

Ultraluminous supersoft sources, with bolometric luminosities $\ge 10^{39}$ erg
s$^{-1}$ and supersoft spectra of $50-100$ eV, are more promising IMBH
candidates (Kong et al. 2004, 2005).
As compared to ULXs, ULSSSs exhibit spectra that do not show a hard power-law
component, making it (more) reasonable to link the temperature to the inner
edge of the accretion disk, thereby strongly suggesting an IMBH as the
accretor.
ULSSSs can be the ultraluminous version of the canonical supersoft sources as
ULXs can be stellar black holes.  However, given the  nature and dimension of
the surface nuclear burning for white dwarfs,  it is unlikely to make apparent
super-Eddington emitters with beaming effects or the photon-bubble instability
in magnetized accretion disks as proposed for stellar black holes in ULXs.
Of course, applying white dwarf atmosphere models to supersoft source spectra
may reduce their bolometric luminosities by a factor of ten (van Teeseling et
al. 1996), hence alleviating the super-Eddington problem for some ULSSSs, but it
is unlikely to push the most luminous ones below the Eddington limit.

Swartz et al. (2002) observed the nearby spiral galaxy M81 with Chandra ACIS,
and discovered an intriguing ULSSS in the bulge at R.A.=09:55:42.2,
Decl.=69:03:36.5, which we designate as M81-ULS1 in this paper.
Liu (2007) studied the spectral properties of this ULSSS with 17 Chandra ACIS
observations, and found its spectra persistently supersoft in six years despite
dramatic flux changes.
The spectrum from the longest observation 
can be well fitted in 0.3-2 keV by an absorbed blackbody model for a nuclear
burning white dwarf, with $\chi^2_\nu/dof = 1.196/38$, $n_H =
8.6\pm0.9\times10^{20}$ cm$^{-2}$, $kT = 73\pm1.5$ eV, $L_X(0.3-2keV) =
3.2\times10^{38}$ erg s$^{-1}$, and $L_{bol} = 2.5\times10^{39}$ erg s$^{-1}$.
%
The spectrum can be equally well fitted by a multicolor disk  model, with
$\chi^2_\nu/dof = 1.073/38$, $n_H = 10.5\pm0.9\times10^{20}$ cm$^{-2}$,
$kT_{in} = 83\pm1.4$ eV and $L_X(0.3-2keV) = 3.2\times10^{38}$ erg s$^{-1}$.
The inner disk radius derived from the model normalization is $R_{in} \approx
24500/(cosi)^{1/2}$ km, which corresponds to an IMBH of $\sim2700/(cosi)^{1/2}
M_\odot$, assuming $R_{in}$ corresponds to the last stable orbit. 
Thus, the X-ray spectra are suggestive of M81-ULS1 being either a white dwarf or an
IMBH.

Observations in other wavelengths can shed light on the nature of this class of
X-ray sources. We have embarked on an effort to study such sources in nearby
galaxies with HST observations.  In this letter we report the optical studies
of M81-ULS1 with HST archive data. \S2 describes the HST archive data sets
utilized, the optical identification and photometric measurements of ULS1. The
photometric measurements of ULS1 are interpreted in light of the white dwarf
model and the IMBH model in \S3. We discuss our results and their implications
in \S4.  The distance of M81 is taken to be 3.63 Mpc ($\mu=27.8$ mag; Freedman
et al. 1994) in this work.

\section{ANALYSIS OF HST DATA}

HST has observed the sky region around ULS1 several times (Table 1). These
include an observation on 1996-01-01 in filters WFPC2/F547M (3$\times$30s) and
WFPC2/F656N (3$\times$600s), an observation on 2003-09-18 in filters ACS/WFC
F658N (700s) and ACS/WFC F814W (120s), an observation on 2004-09-14 in ACS/WFC
F814W (1650s), and an observation on 2006-03-27 in filters ACS/WFC F435W (465s)
and ACS/WFC F606W (470s).  All data sets were downloaded from MAST and
calibrated on-the-fly with the best calibration files as of November 7, 2007.


To identify the X-ray sources on the HST images, we register both the HST
images and the Chandra images onto the 2MASS reference frame. 
This was achieved by identifying 11 2MASS sources on the ACS/WFC F814W image
taken in 2004, and two 2MASS sources on the combined Chandra image.
After both were registered onto the 2MASS reference frame, the X-ray image was
overlayed onto the ACS/WFC F814W image. 
Six X-ray sources were coincident with optical objects within error circles of
$0\farcs5$, including ULS1 that is identified with a point-like object. 
This object appeared in all HST observations (Figure 1), and is located in the
smooth bulge of M81 with the closet object of equal brightness in F814W
$10^{\prime\prime}$ away.


The photometry for this point-like object was calculated with an aperture of
$0\farcs5$  on WFPC2/PC1 and ACS/WFC images.  
The count rates and the corresponding magnitudes in the vegamag photometric
system are listed in Table 1 for the counterpart in the seven observations.  
We calculate the magnitudes for two $H_\alpha$ filters with SYNPHOT assuming an
$H_\alpha$ emission line atop the continuum of the Bruzual stellar spectrum
bz\_55; this spectrum, with an intrinsic color $F547M - F814W \approx 1.3$ as
similar to observed, is normalized to give the observed F547M magnitude.
%
The observations require an $H_\alpha$ emission line luminosity of
$(1.9-3.2)\times10^{36}$ erg s$^{-1}$ during the 1996 WFPC2/F656N observation,
and $(0.8-1.1)\times10^{36}$ erg s$^{-1}$ during the 2003 WFC/F658N
observation.

\section{INTERPRETATION OF HST OBSERVATIONS}

The optical counterpart of ULS1, placed at the distance of M81, has an absolute
magnitude of $M_{F814W} \approx -7$ mag, a color of $F547M - F814W \approx 1.3$
as for K-M stars, and $F435W - F547M \approx -0.1$ as for late B stars.
The red color and absolute magnitudes in F547M and F814W might be obtained from
a massive old globular cluster of $\sim10^6$ K-M dwarfs. However, the same
globular cluster can not explain the late B-like blue color of $F435W - F547M
\approx -0.1$.  Furthermore, the variabilities in the $H_\alpha$ and F814W
bands suggest that the counterpart is a single object.
For single objects, such a brightness and red color in F547M and F814W are not
seen in dwarf/giant/supergiant stars as tabulated in {\it Allen's Astrophysical
Quantities} (Cox 2000), but can be seen in asymptotic giant branch (AGB) stars
that are burning hydrogen/helium in shells and shine as a blackbody.
We note, however, the object is considerably bluer than AGB stars with the same
F814W luminosities, and there must be a bluer component in addition to the AGB
star to account for the observed colors and magnitudes simultaneously. 
To quantify the AGB star properties in term of the core mass ($M_c$), we adopt
the luminosity core-mass relation $L = 59250 (M_c - 0.522)$ (Paczynski 1970),
the core-mass radius relation $R/R_\odot = R_0 M_c^{4.5}/(1+4M_c^4)+0.5$ ($R_0
= 4950$; Rappaport et al. 1995) and Stefan-Boltzman's law $L=4\pi R^2\sigma
T^4$.

A natural explanation for the blue component is that it originates from the
accretion disk.
If the accretor is a white dwarf, the accretion disk is illuminated by the soft
X-ray emission from the white dwarf surface.
Given the high X-ray luminosity, such a disk shall flare up geometrically,
intercept a large fraction of the X-ray emission and re-emit in lower energies
(Popham \& Di Stefano 1996).
The emergent spectrum will take a form of $\nu^{-1}$ in the optical (FKR 2002).
We calculate the F547M, F435W, 606W, and F814W (2004) magnitudes with SYNPHOT for an
AGB star plus a powerlaw of $\nu^{-1}$, for different core masses and relative
contributions from the two components, and compute the quantity $\Delta m^2
\equiv \Sigma (m_{obs} - m_{model})^2$ as a sum of the squared differences
between the observed and model magnitudes.
By minimizing $\Delta m^2$, we found at $\Delta m^2 = 0.4$ the best combination
of the $\nu^{-1}$ powerlaw component and an AGB star, which has 
$M_c = 0.77 M_\odot$, $R = 639 R_\odot$ and $T = 2444 K$.
This model, as plotted in Figure 2, can not explain the observations because it
does not give enough blue light in F435W. 
The blackbody spectral fit to the X-ray data (Liu 2007) is overplotted for
comparison; clearly, a white dwarf itself would be too dim in the optical to
explain the observations.

If the accretor is an IMBH and thus the high X-ray luminosity is less than a
few percents of the Eddington luminosity, the accretion disc will be a standard
geometrically-thin optically-thick one (Shafee et al. 2007).  
The spectrum of such a thermal disk is conventionally described by a
multi-color disc (MCD) model (Mitsuda et al. 1984).
The power-law component with $F_\nu \propto \nu^{1/3}$ for the MCD spectrum may
extend down to the optical if the outer disk radius is large.  For 
ULS1, this means $R_{out} \sim 10^2 R_{in}$, which is reasonably small ($\sim
40/(cosi)^{1/2} R_\odot$) as compared to the giant star size or the Roche lobe
size of the black hole. 
With similar procedures as in the white dwarf model, the four observed
magnitudes determine a best combination, at $\Delta m^2 = 0.16$, of the
$\nu^{1/3}$ powerlaw component and an AGB star, which has $M_c = 1.30 M_\odot$,
$R = 1301 R_\odot$ and $T = 2277 K$.
The model, plotted in Figure 3 with the MCD model fit (Liu 2007) overplotted
for comparison, can explain all four observed magnitudes.
Remarkably, the required $\nu^{1/3}$ power-law component in the optical is
consistent to within 30\% with the MCD fit to the X-ray spectrum.


The properties of the companion AGB star may differ considering the
uncertainties in the theory and the observations.
In particular, the analytical form for the core-mass radius relation for AGB
stars, adopted for the convenience of calculations, tends to overestimate the
radius (Rappaport et al. 1995).
To test how this affects our results, we re-calculate the models by setting
$R_0 = 3300$, i.e., reducing the radius by 33\%.  
Under this simplistic modification, we find an AGB star with $M_c = 0.60
M_\odot$, $R = 219 R_\odot$ and $T = 3119 K$ for the white dwarf model, and an
AGB star with $M_c = 0.93 M_\odot$, $R =  595 R_\odot$ and $T = 2859 K$ for the
IMBH model.
The white dwarf model has a $\Delta m^2 = 0.42$ and cannot explain all the
observations because it does not give enough blue light.
The IMBH model can fit all the observations simultaneously, but to a less
satisfactory level ($\Delta m^2 = 0.26$) as compared to the model with $R_0 =
4950$ ($\Delta m^2 = 0.16$).
The newly computed AGB star is hotter and more compact. It has a less massive
core ($0.93 M_\odot$), which is consistent with the final core masses of AGB
stars with heavy stellar wind (Weidemann 2000).
If the AGB star is at its final stage, its core mass indicates an initial mass
greater than $5 M_\odot$ and a lifetime shorter than 0.1 Gyrs (Vassiliadis \&
Wood 1993).

\section{DISCUSSION}



Ultraluminous supersoft sources (ULSSSs), as X-ray spectroscopy suggests, are
possibly the long-sought intermediate mass black holes (IMBHs), or massive
white dwarfs that are progenitors of type Ia supernovae.
%
%
We note that the accretion disk behaves quite differently in presence of an
IMBH accretor or a white dwarf accretor for the same X-ray luminosities, and
observations in optical and infrared are capable to pick up the differences.
In this Letter, we report the optical studies with HST archive observations of
M81-ULS1, a ULSSS in the M81 bulge.
To summarize, the spectral energy distribution of M81-ULS1 in the optical
reveals a blue component in addition to the companion of an AGB star.  
The blue component is consistent with a $F_\nu \propto \nu^{1/3}$ power-law as
expected from the geometrically-thin accretion disk around an IMBH accretor,
but inconsistent with a $F_\nu\propto\nu^{-1}$ pow-law as expected from the
X-ray irradiated flared acrretion disk around a white dwarf accretor.
This is strong observational evidence that M81-ULS1 is an IMBH instead of a
white dwarf.
%
%

The HST archive observations, however, were taken at different times spanning
six years.
This may affect our conclusion because the differences between different bands
could result (partly) from the intrinsic variabilities of the accretion disk
and the AGB companion.
The accretion disk may have changed as suggested by the X-ray variability over
the years, while AGB stars are known to be long period variables with
amplitudes of up to $\Delta V \sim 2.5$. 
The marginally significant change in the F814W magnitudes from two observations
separated by one year may have reflected such variabilities.
Simultaneous observations in multiple bands are thus required to make clear
whether and how much the spectral energy distribution in this work was
contaminated by the variabilities.
While such simultaneous observations may change the details of the AGB
companion and the accretion disk, they probably will not change the main
conclusion that the blue component comes from a $F_\nu\propto\nu^{1/3}$
powerlaw instead of a $F_\nu\propto\nu^{-1}$ power. 
This is because the distinguishing power of the optical data mainly comes from
the blue color between F435W and F606W (or F547M) as illustrated in Figures 2
and 3, yet F435W and F606W were already observed simultaneously.

The conclusion would be invalidated if the optical counterpart were an
interloper close to M81-ULS1 by projection.
Such projection, however, is unlikely because the probability of finding such
an AGB star in a $0\farcs5$ error circle by chance is quite small ($<0.003$),
as calculated from the the error circle size and the fact that no stars of
equal brightness were found within $10\farcs$.
Also, the prominent $H_\alpha$ emission is suggestive of the photoionization of
surrounding materials by the soft X-ray emission from ULS1, and the $H_\alpha$
variation between two observations is somehow expected due to the long term
X-ray variability of ULS1 (Liu 2007).
Furthermore, a companion of an AGB star with a supergiant size, if overflowing
or underfilling its Roche lobe, implies a period of 30 years or longer, which
is consistent with the failure to detect any period from the six year Chandra
observations (Liu 2007).
%


\acknowledgements

We would like to thank Douglas Swartz, Scott Kenyan and Jeff McClintock for
helpful discussions.  JFL acknowledges the support for this work provided by
NASA through the Chandra Fellowship Program, grant PF6-70043. RD acknowledges
the supports from grants NAG5-10705, AR-10948.019 and NNX07AH64G-R.

\clearpage


\begin{figure}

\plotone{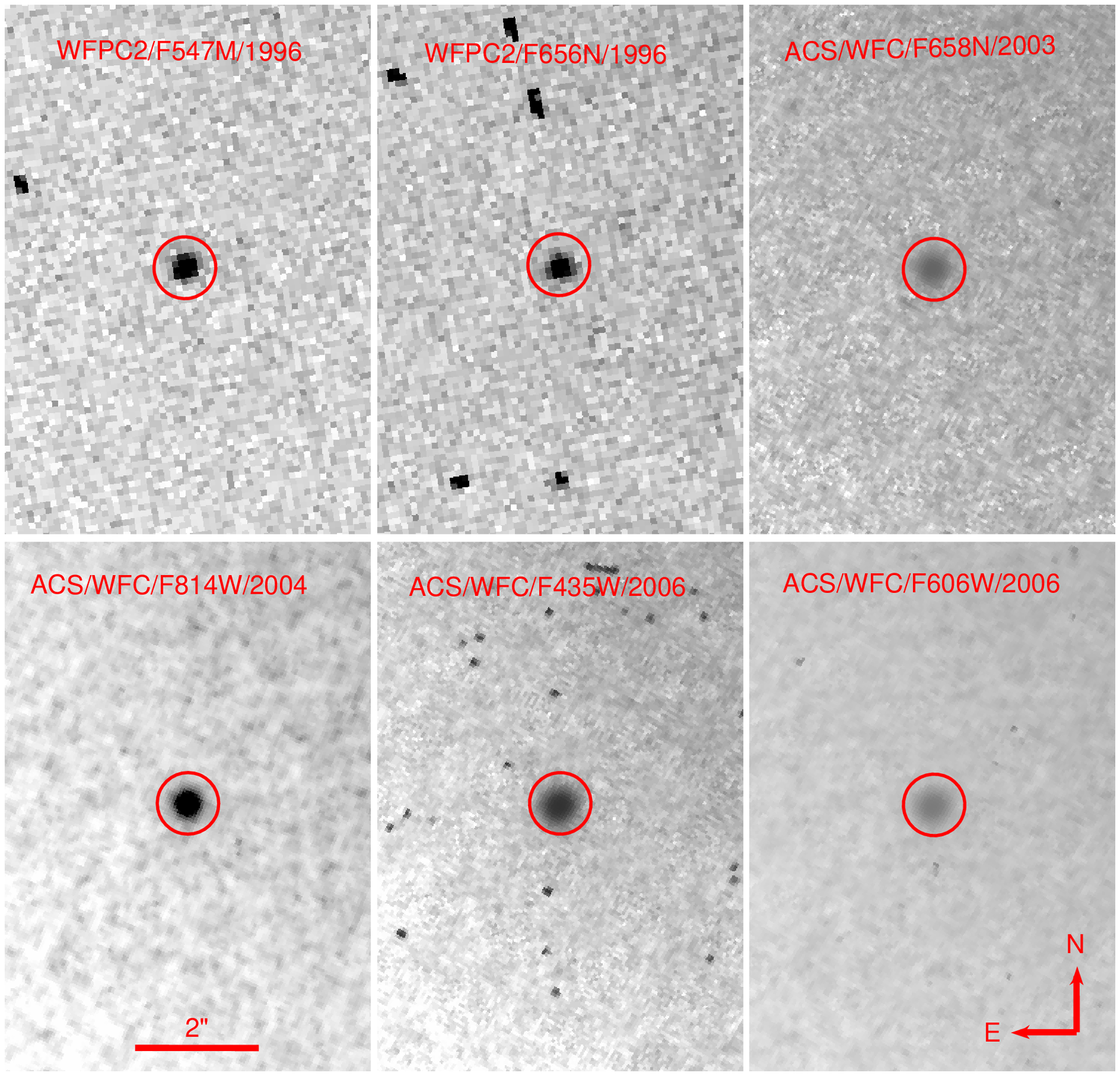}

\caption{The optical counterpart for M81-ULS1 on HST images in F547M (1996),
F656N (1996), F658N (2003), F814W (2004), F435W (2006) and F606W (2006).  The
WFPC2 F547M/F656N and ACS/WFC F435W images are plagued with cosmic rays.  All
images are aligned with the counterpart at the image centers. The error circles
have a radius of $0\farcs5$.  North is up in all images. The counterpart also
appeared on the ACS/WFC F814W (2003) image, which was very noisy due to the
short exposure and not shown here. }

\end{figure}

\begin{figure}

\plotone{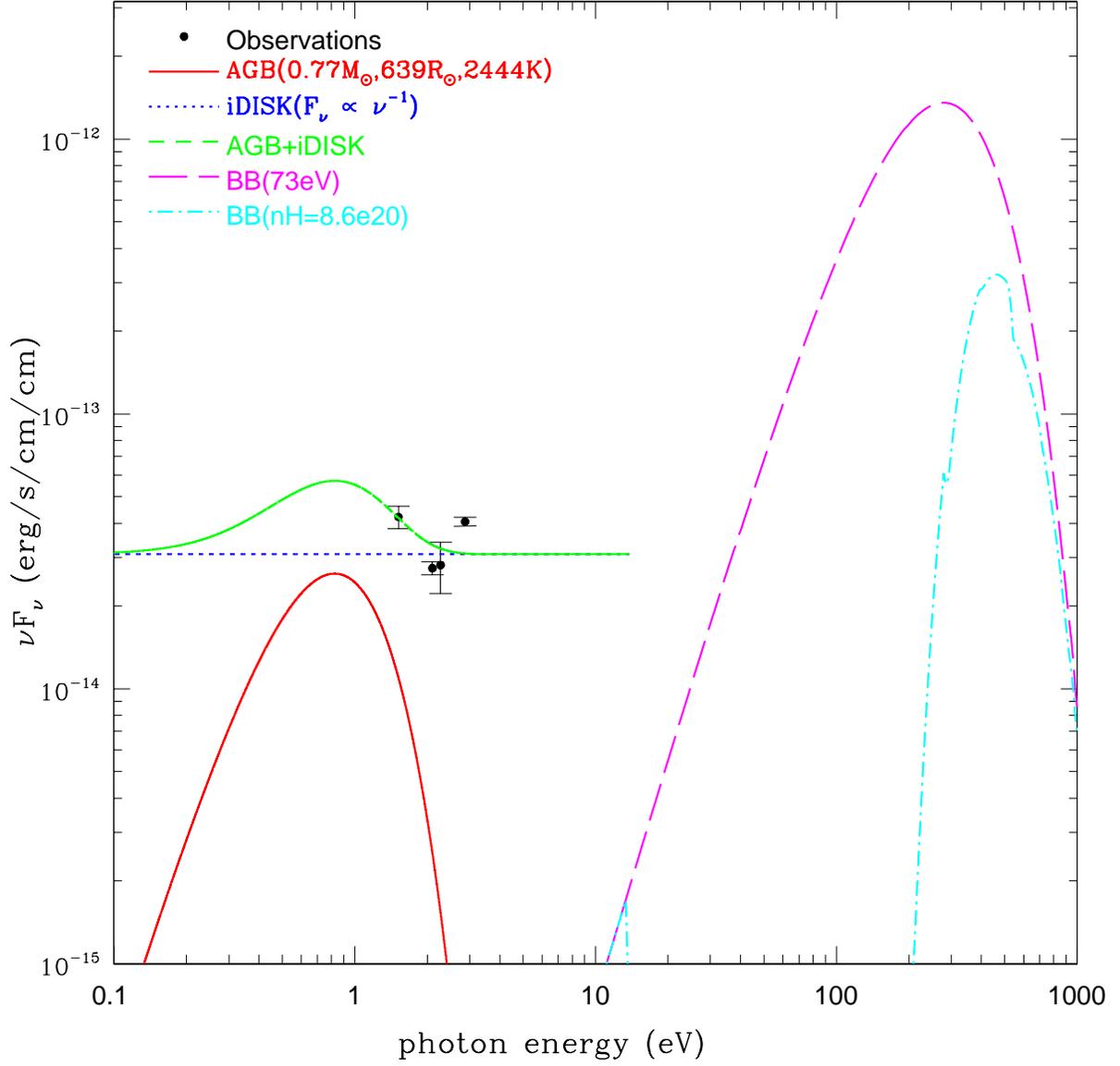}

\caption{Fit for the optical observations with an AGB star plus an accretion
disk illuminated by X-ray emission from the surface of a nuclear burning white
dwarf. The X-ray illuminated accretion disk radiates as a power of
$F_\nu\propto\nu^{-1}$.  The unabsorbed and absorbed spectral fits as a 73 eV
blackbody for the white dwarf X-ray emission are plotted for comparison.  }

\end{figure}

\begin{figure}

\plotone{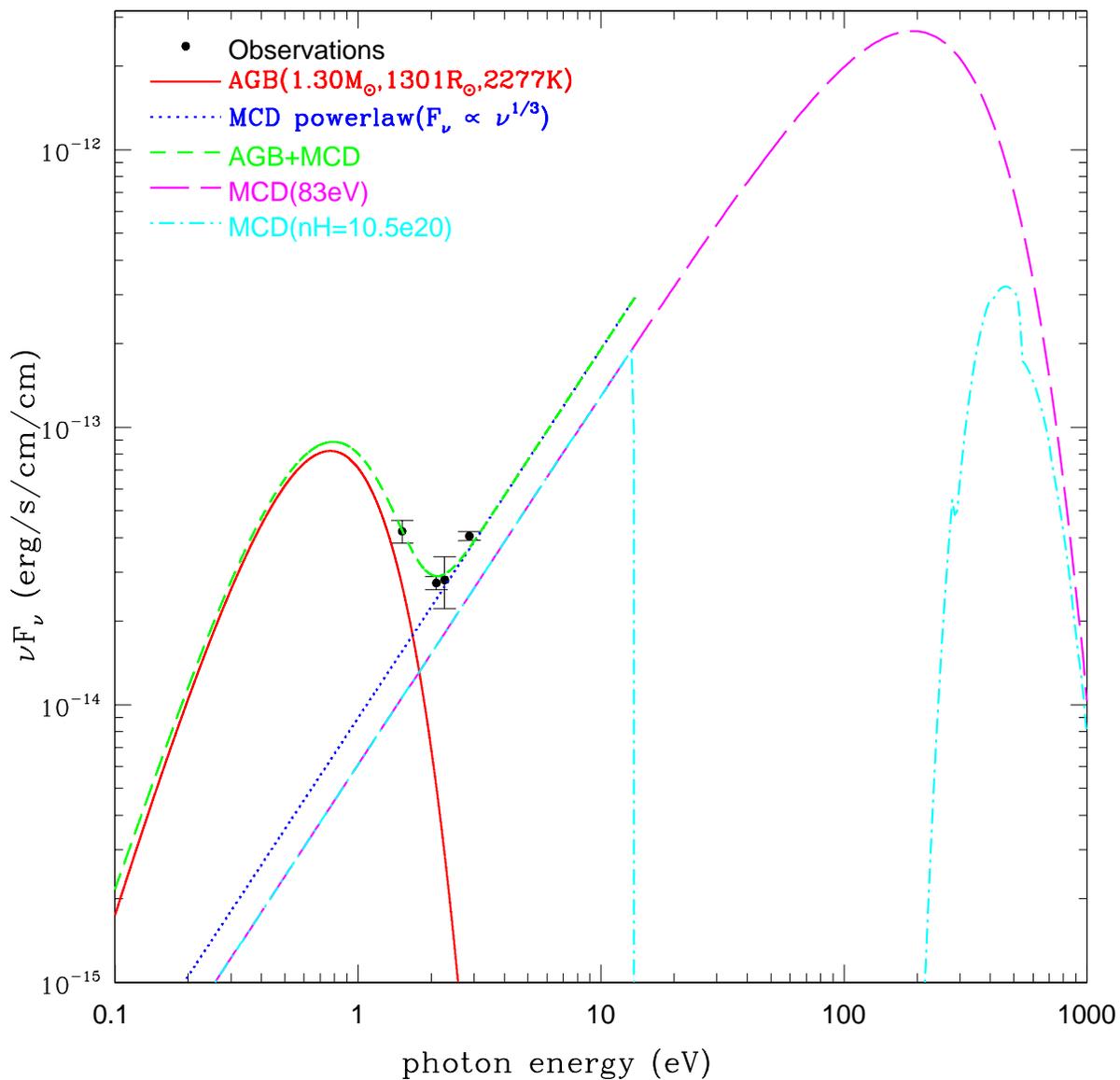}

\caption{Fit for the optical observations with an AGB plus the power-law
component of a multicolor disk around an IMBH. The power-law component has a
form of $F_\nu\propto\nu^{1/3}$. The multicolor disk has an inner edge
temperature of 83 eV as from X-ray analysis, and its unabsorbed and absorbed
spectra are plotted for comparison. }

\end{figure}



\begin{deluxetable}{llllll}
\tablecaption{HST observations for M81-ULS1}
\tablehead{
\colhead{ID} & \colhead{Filter} & \colhead{ExpT} & \colhead{DATE} & \colhead{Count Rate} & \colhead{ VEGAMAG } \\
}

\startdata
U32L0102/5/8 & WFPC2/F547M & 3x30  & 1996-01-01 & $0.67 \pm 0.17$	& $22.1\pm0.2$  \\
U32L0101/4/7 & WFPC2/F656N & 3x600 & 1996-01-01 & $0.09 \pm 0.01$	& $20.2 \pm 0.1$  \\
J8MX18010  & WFC/F658N   & 700   & 2003-09-18 & $8.7 \pm 0.5$	& $20.0 \pm 0.05$  \\
J8MX18AKQ  & WFC/F814W   & 120   & 2003-09-18 & $82.4 \pm 5$	& $20.7 \pm 0.06$  \\
J90LA5010  & WFC/F814W   & 1650  & 2004-09-14 & $73.3 \pm 3$	& $20.8 \pm 0.04$  \\
J9EL28A9Q  & WFC/F435W   & 465   & 2006-03-27 & $29.9 \pm 0.5$	& $22.09 \pm 0.02$  \\
J9EL28AAQ  & WFC/F606W   & 470   & 2006-03-27 & $58.7 \pm 1.7$	& $21.98 \pm 0.03$  \\

\enddata

\tablecomments{The columns are (1) exposure ID, (2) filter, (3) total exposure
in seconds, (4) observation date, (5) count rate in e$^-$/s for the
counterpart, and (6) magnitudes in the VEGAMAG photometric system. }

\end{deluxetable}

\end{document}